\documentclass[useAMS,usenatbib]{mn2e}
\usepackage{graphicx}
\usepackage{url}
\usepackage{aas_macros}
\usepackage{amssymb}

\title[The variability plane]{The variability plane of accreting compact objects}
\author[E. G. K\"ording, S. Migliari, R. Fender, T. Belloni, C. Knigge, and I. M$^{\mathrm c}$Hardy]{E. G. K\"ording$^{1}$\thanks{E-mail:
Elmar@phys.soton.ac.uk}, S. Migliari$^2$, R. Fender$^{1}$, T. Belloni$^3,$ C. Knigge$^{1}$, and I. M$^{\mathrm c}$Hardy$^{1}$\\
$^{1}$ School of Physics and Astronomy, University of Southampton Hampshire SO17 1BJ, United Kingdom \\
$^{2}$ Center for Astrophysics and Space Sciences, Code 0424, University of California at San Diego, La Jolla, CA 92093, USA \\
$^{3}$ INAF-Osservatorio Astronomico di Brera, Via E. Bianchi 46, I-23807 Merate (LC), Italy\\
}

\begin{document}

\date{Accepted ? Received ? }

\pagerange{\pageref{firstpage}--\pageref{lastpage}} \pubyear{2007}

\maketitle

\label{firstpage}

\begin{abstract} 
Recently, it has been shown that soft-state black hole X-ray binaries and active galactic nuclei populate a plane in the space defined by the black hole mass, accretion rate and characteristic frequency. 
We show that this plane can be extended to hard-state objects if one allows a constant offset for the frequencies in the soft and the hard state. During a state transition the frequencies rapidly move from one scaling to the other depending on an additional parameter, possibly the disk-fraction. The relationship between frequency, mass and accretion rate can be further extended by including weakly accreting neutron stars. We explore if the lower kHz QPOs of neutron stars and the dwarf nova oscillations of white dwarfs can be included as well and discuss the physical implications of the found correlation.
\end{abstract}


\section{Introduction}
The general idea that X-ray binaries (XRBs) and active galactic nuclei (AGN) have similar central engines, if scaled with black hole mass, is increasingly supported by empirical correlations connecting both classes. First, the fundamental plane of accreting black holes connects XRBs and AGN through a plane in the black hole mass, radio and X-ray luminosity space \citep{MerloniHeinzdiMatteo2003,FalckeKoerdingMarkoff2004}. Recently, \citet{McHardyKoerding2006} have reported that a second plane connecting stellar and supermassive black holes exists in the space defined by the accretion rate, the black hole mass and a characteristic timescale of the X-ray variability. This suggests that all black holes can be unified by taking the accretion rate as well as the black hole mass into account. Here, we explore the latter correlation in further detail and discuss the inclusion of strongly sub-Eddington black holes (BHs) and neutron stars (NSs).

\begin{figure}
\resizebox{8.7cm}{!}{\includegraphics{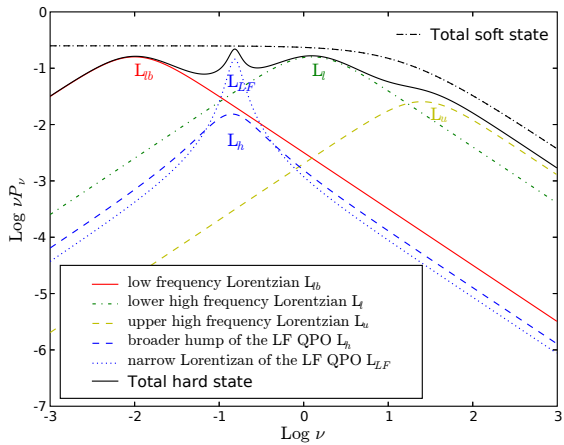}}
\caption{Sketch of the different Lorentzians found in BH and NS XRBs. The $x$-axis denotes the frequency and on the $y$-axis we show frequency times power. Besides the Lorentzians found in hard-state objects we also show a broken power law model found for the PSDs of soft-state BHs. We have increased the normalisation of the soft-state model so that it lies above the model for the hard state.}
\label{fiSketch}
\end{figure}

The power spectral density (PSD) of NS and BH XRBs can be well described by a number of Lorentzians with variable coherence factor Q \citep[e.g.,][]{PsaltisBelloni1999,BelloniPsaltis2002}. Each Lorentzian is described by their characteristic frequency (the frequency of the maximum of the Lorentzian in the frequency times power plot). As presented in \citet{BelloniPsaltis2002}, it is possible to fit NSs and BHs with  
\begin{itemize}
\item a zero-centered low frequency Lorentzian $L_{\mathrm lb}$ fitting the low-frequency end of the band-limited noise with characteristic frequency $\nu_{\mathrm lb}$. Usually, this frequency is denoted as $\nu_b$ in the literature. However, to avoid confusion with the break frequency studied in AGN, we add the prefix "l" for low.
\item two Lorentzians fitting the high-frequency end of the band-limited noise with frequencies $\nu_l$ and $\nu_u$ (the lower and upper high-frequency Lorentzian $L_l$ and $L_u$). These two Lorentzians take over the role of the upper and lower kHz QPO if they are present.
\item one or two Lorentzians fitting the region around the frequency of the low-frequency quasi-periodic oscillation (QPO). The narrow core of the QPO ($L_{LF}$) has the characteristic frequency $\nu_{LF}$, while the broader "hump" Lorentzian $L_h$ has a characteristic frequency $\nu_h$.  
\end{itemize}
For a sketch showing the different Lorentizans see Fig.~\ref{fiSketch}.

XRBs are observed in several states: the hard, the soft and the intermediate state (IMS). For definitions, subclasses and examples see e.g., \citet{Nowak1995,BelloniHomanCasella2005,HomanBelloni2005}, but see also \citet{McClintockRemillard2003} for slightly different definitions.

The measured shapes of AGN PSDs are similar to those of soft-state XRBs (see e.g, \citealt{McHardy1988,MarkowitzEdelsonVaughan2003}). The timescales for AGN are several orders of magnitudes longer, roughly in agreement with the expectation that all length-scales increase linearly with black hole mass. The uncertainties of the measured AGN PSDs are usually too large to fit a detailed Lorentzian model (but see McHardy et al. in prep.), due to the long timescales. Thus, one uses broken power law models. The shape of the PSD of AGN and XRBs at high Fourier frequencies is well described by a power law with index $\sim -2$ (e.g., \citealt{GreenMcHardyLehto1993}, \citealt{CuiZhangFocke1997}).  For Seyfert Galaxies and those soft-state XRBs similar to Cyg~X-1 the PSD at low Fourier frequencies turns to a power law with index $\sim -1$ (see Fig.~\ref{fiSketch}, \citealt{EdelsonNandra1999,UttleyMcHardyPapadakis2002,CuiZhangFocke1997}).  The Fourier frequency of the turnover will be denoted as the high-break frequency $\nu_{\mathrm hb}$. Using the bolometric luminosity as a tracer for the accretion rate, \citet{McHardyKoerding2006} found that this high-break frequency is related to the accretion rate and black hole mass as $\nu_{\mathrm hb} M \propto \dot{M}/\dot{M}_{\mathrm Edd}$ for soft-state objects.

It has recently been found by \citet{MigliariFender2005} that for the BH GX 339-4 the frequency of the lower high-frequency Lorentzian $\nu_l$ and for the NS 4U~1728-34 the frequency of the "hump"-Lorentzian $\nu_h$ is correlated with the radio luminosity in their hard state as $\nu_{l,h} \propto L_R^{0.7}$, where $\nu_{l,h}$ denotes either $\nu_l$ or $\nu_h$. 
As the radio luminosity is a good tracer of the accretion rate (\citealt{KoerdingFenderMigliari2006}, $L_R \propto \dot{M}^{1.4}$), the correlation can be written as $\nu_{l,h} \propto \dot{M}$. It is therefore suggestive that the "variability plane" found for high state XRBs and AGN can be extended to hard-state BHs or even NSs. 

As shown in Fig.~\ref{fiSketch} the superposition of the different Lorentzians ($L_{lb}$, $L_h$, $L_l$ and $L_u$) creates a flat plateau in the $\nu P_\nu$ plots, and the PSD can therefore be roughly described with power law with index $\sim -1$ for frequencies between $\nu_{lb}$ and $\nu_l$. This power law steepens to a power law with index $\sim -2$ at the peak of the high-frequency Lorentzians ($L_l$ and $L_u$) similar to the behaviour found in soft-state XRBs and AGN. It is therefore likely, that the frequency of either the upper or lower high-frequency Lorentzian ($\nu_u$ or $\nu_l$) correspond to the high break frequency found in soft-state XRBs and AGN. 
The lower high-frequency Lorentzian $L_l$ usually dominates the high-frequency end of the band limited noise \citep[e.g.,][]{Pottschmidt2002}. Additionally, $\nu_l$ and $\nu_u$ are not linearly coupled \citep[e.g.,][]{BelloniPsaltis2002}. While the frequency of the lower high-frequency Lorentizan $\nu_l$ seems to be proportional to $\dot{M}$, $\nu_u$ is therefore not linearly proportional to $\dot{M}$ for individual sources. This suggests, that the lower high-frequency Lorentzian $L_l$ with the characteristic frequency $\nu_l$ corresponds to the high break frequency of AGN.

We note that for NS one observes "parallel tracks" for kHz QPOs on frequency flux diagrams \citep[e.g.,][]{Ford2000,Klis2001}, ie., while flux and frequencies are strongly correlated for short periods of time, the correlation is nearly absent for longer timescales.  
 If the X-ray luminosity (flux) is a good tracer of the accretion rate, there 
cannot be a simple one to one dependence of the frequency $\nu_l$ of the lower high-frequency Lorentzian on accretion rate. However, there are at least three possible tracers of the accretion rate (frequencies, X-ray luminosity and radio luminosity) and the frequencies or the radio luminosity might be better indicators than the X-ray luminosity.

\citet{PsaltisBelloni1999,BelloniPsaltis2002} present a correlation of the frequencies of the lower kHz QPO with those of the low frequency QPO $\nu_{LF}$. This correlation can be extended to include noise features of BHXRBs by using the lower high-frequency Lorentzian $L_l$ as the lower kHz QPO. Also cataclysmic variables (CVs) can be included by using dwarf nova oscillations (DNOs, e.g., \citealt{WarnerWoudtPretorius2003}) and the QPO frequency. Here, we explore if the varability plane found in sort state AGN and XRBs can also be extended to hard-state BH XRBs and NSs or even to white dwarfs.

\section{The sample}
To explore the dependence of the variability properties on the accretion rate and mass of the compact object, we construct a sample of hard and soft-state BHs and NSs with estimated accretion rates. 

\subsection{Accretion rates} \label{seacc}
Soft state BHs and all NSs are generally assumed to be efficiently accreting. Thus, we can use the bolometric luminosity directly as a measure of the accretion rate:
\begin{equation}
\dot{M} = \frac{L_{bol}}{0.1 c^2}, 
\end{equation}
where we assumed an accretion efficiency of $\eta = 0.1$.

For hard-state BHs the accretion rate is not linearly related to the X-ray as the accretion flow is likely to be inefficient \citep[e.g.,][]{EsinMcClintockNarayan1997}.
However, \citet{KoerdingFenderMigliari2006} provide estimates of the accretion rates from either radio or X-ray luminosities. The accretion measure based on the radio luminosity is
\begin{equation}
\dot{M} = 3 \times 10^{17}  \left(\frac{L_{\mathrm Rad}}{10^{30} \mbox{erg s}^{-1}} \right)^{12/17} \frac{\mbox{g}}{\mbox{s}}.
\end{equation}
In the notation given in \citet{KoerdingFenderMigliari2006} we have set $f=1$ and $\eta = 0.1$ to allow for a direct comparison between hard-state BHs and sources with a measured accretion rate obtained from the bolometric luminosity.
Using the fundamental plane of accreting black holes (and the radio/X-ray correlation for XRBs) the radio luminosity accretion measure can be translated to an accretion rate estimate based on the 2-10 keV X-ray luminosity:
\begin{equation}
\dot{M} \approx 3.4 \times 10^{17} \left(\frac{L_{2-10keV}}{10^{36} \mbox{erg s}^{-1}} \right)^{0.5} \left(\frac{M}{M_{GX339}} \right)^{0.43} \frac{\mbox{g}}{\mbox{s}}.
\end{equation}
We note, that this accretion measure does not only depend on the X-ray luminosity but also on the mass of the black hole. While the masses of BH XRBs are typically around $\sim 10 M_\odot$ this mass-term provides an additional uncertainty compared to the accretion rate measure based on the radio luminosity. Thus, we will estimate the accretion rate from the radio luminosity if quasi-simultaneous measurements of the radio flux and the timing features are available. Otherwise, we have to use the X-ray flux to obtain an accretion rate estimate. 

The uncertainty of these accretion rate measures is hard to access as there are only a few data-points available to normalise the accretion measure. The sample standard-deviation of the 4 BH data-points is $\sim 0.3$ dex, but has little significance. If one includes also the NS points (overall 14 points) we find a scatter of $\sim 0.2$ dex. 
Another possibility to measure the uncertainty of the accretion measure is via the fundamental plane of accreting BHs. If both accretion measures (based on the radio and X-ray emission) presented here are exact, there would be no scatter in radio/X-ray correlation found for XRBs and the fundamental plane for XRBs and AGN. The scatter around the fundamental plane can therefore be used as rough estimator of the uncertainty of the accretion rate measure. The cleanest sample for the fundamental plane including only hard-state XRBs and low luminosity AGN has a scatter of 0.15 dex \citep{KoerdingFalckeCorbel2005}. This suggests that the uncertainties of the accretion rate measure based on radio luminosity is around $\sim 0.2$ dex. The accretion rate measure based on the X-ray luminosity will have a similar intrinsic uncertainty, but has additional uncertainties due to uncertainties of the BH mass estimates. 

The accretion rate measure based on the radio luminosity is also applicable to island state NS. However, as the radio luminosity accretion rate measure has been normalised using accretion rates obtained from bolometric luminosities its absolute accuracy is unlikely to exceed the measure based on bolometric luminosities for efficiently accretion objects. In summary we will use the following accretion rate measures in order of preference:
\begin{enumerate}
\item Accretion rates from bolometric luminosity for soft-state BHs and NSs
\item Accretion rates estimated from the radio luminosity for hard-state BHs
\item Accretion rates estimated from the X-ray luminosity for hard-state BHs
\end{enumerate}

Throughout this paper we will measure the accretion rate in g/s, all frequencies in Hz and masses of compact objects in solar masses. 

\subsection{Black Holes}
Our hard-state BH sample is based on measurements in the public RXTE archive as well as already published data.
We select all sources which have well measurable frequencies $\nu_l$ as well as estimates for the black hole mass and distance. However, in order to obtain an estimate of $\nu_l$ one needs a hard-state power spectrum, since in the hard-intermediate state characteristic frequencies are higher and the broad lower high-frequency Lorentzian ($L_l$) cannot usually be detected (with the exception of GRS 1915+105). Moreover, the source count rate needs to be sufficiently high to allow a detection. This rules out the final parts of the outbursts and limits our sample to bright early hard-state observations. Unfortunately, only few sources have been observed sufficiently early in their outburst to fulfill all the requirements:

\begin{itemize}
\item {\it GX~339-4}: \citet{MigliariFender2005} present values for $\nu_l$ and the radio fluxes of GX~339-4 in its hard state. These radio fluxes have been used to estimate the accretion rate. The distance to GX~339-4 is still uncertain,  \citet{ShahbazFenderCharles2001} and \citet{JonkerNelemans2004} give a lower limit of 6 kpc, but the distance may be as high as 15 kpc \citep{HynesSteeghsCasares2004}. We therefore adopt a distance of $8$ kpc. 
The mass function is $5.8 \pm 0.5$ $M_\odot$ \citep{HynesSteeghsCasares2003}, which is therefore a lower limit for the mass of the black hole. If one assumes zero mass for the companion and a mean inclination angle we obtain a mass of  $M \approx 12 M_\odot$. Additionally, this higher mass fits the fundamental plane as well as timing correlations better than smaller masses. Thus, we use $M = 12 M_\odot$.

\item {\it XTE J1118+480}: \citet{BelloniPsaltis2002} measured $\nu_l$ for XTE J1118+480 on the 4th and 15th May 2000. The VLA has observed the
source during that time. We have analysed archival 8.5 GHz data and found a flux of 5.7 mJy on the 27th April,  6.4 mJy on the 13th May, and 6.8 mJy on the 31st May. We interpolate linearly between those dates and estimate the 8.5 GHz flux on the 4th of May to be 5.9 mJy and 6.4 mJy on the 15th May and estimate the accretion rate from these radio fluxes. We assume a mass of $6.8 \pm 0.3 M_\odot$ \citep{RitterKolb2003} and a distance of $1.71 \pm 0.05$ kpc \citep{ChatyHaswellMalzac2003}.

\item {\it XTE J1550-564}: For the outburst of XTE J1550-564 in 2002 we use $\nu_l$ values and X-ray fluxes from \citet*{BelloniColomboHoman2002}. This outburst never left the hard state. Additionally, we include the outbursts of 1998 and 2000. For these outbursts, we reanalysed the data of the RXTE archive to obtain $\nu_l$. The 2-10 keV X-ray flux has been estimated from the 2 - 9 keV PCA counts assuming that the X-ray spectrum can be described by a power-law with photon index of $\Gamma = 1.5$. As the observed X-ray band is similar to the energy range of the estimated flux, the spectral uncertainties do not strongly effect the estimated flux. We assume a distance of $5.3 \pm 2.3$ kpc \citep{JonkerNelemans2004} and a mass of $10.6 \pm 1$ M$_\odot$ \citep{OroszGroot2002}. 

\item {\it GRO J1655-40}: For GRO J1655-40 we reduced public RXTE data on MJD 53429.7 to measure $\nu_l$ and obtained the 2 - 10 keV flux from the spectral model given on the webpage\footnote{\url{http://tahti.mit.edu/opensource/1655/} } maintained by J. Homan. We assume a mass of $6.3 \pm 0.5 $ $M_\odot$ \citep{GreeneBailynOrosz2001} and a distance of 3.2 kpc \citep{HjellmingRupen1995}, but see \citet{FoellmiDepagneDall2006} for significantly lower distance estimates ($<$1.7 kpc). We therefore assume $3.2 \pm 1.5$ kpc.

\item {\it GS 1354-644}: For GS 1354-644 we assume a mass of $7.34 \pm 0.5$ M$_\odot$ \citep{CasaresZuritaShahbaz2004}. The distance to the source is uncertain,  \citet{KitamotoTsunemiPedersen1990} suggests 10 kpc while  \citet{CasaresZuritaShahbaz2004} give a lower limit of 27 kpc. We will assume a distance of 10 kpc. The 2-10 keV flux and the measurement of $\nu_l$ have been obtained from public RXTE data.

\item {\it XTE J1650-500} RXTE started to observe the source during the outburst in 2001 just as the source starts it transition from the hard to the soft state \citep{HomanKlein-WoltRossi2003}. We use their first observation which is still in the hard state to obtain a value for $\nu_l$ and the 2-10 keV flux. The distance to J1650-500 is between 2--6 kpc \citep{TomsickKalemciCorbel2003}, we assume $4 \pm 2$ kpc. \citet{OroszMcClintockRemillard2004} suggest that the BH mass of J1650-500 is between 4 and 7.3  M$_\odot$, we use a mass of $5.5 \pm 2$ M$\_\odot$. 

\item {\it Cyg~X-1}: To compare these hard-state measurements with a soft-state object, we also show Cyg~X-1 in its soft state. The high break frequencies are taken from \citet{AxelssonBorgonovoLarsson2006}. To obtain accretion rates we use the bolometric luminosities given in \citet{WilmsNowakPottschmidt2006}, as soft-state objects are likely to be efficient accretors. The available data has been binned in luminosity bins as described in \citet{McHardyKoerding2006}. We only include the source for comparison, as the model describing the PSDs for the soft state is different to the Lorentzians we use here. As black hole mass we use $10 M_\odot$, and the distance is assumed to be 2.1 kpc \citep{MasseyJohnsonDegioiaEastwood1995}. 

\item {\it GRS~1915+105}: This source is always accreting at a large fraction of the Eddington accretion rate \citep{FenderBelloni2004}. It is therefore likely to be efficiently accreting, so we can use the bolometric luminosity to estimate the accretion rate \citep{KoerdingFenderMigliari2006}. We obtain $\nu_l$ from \citet{BelloniPsaltis2002} and X-ray fluxes from \citet{Trudolyubov2001}. We assume a mass of 15 $M_\odot$ and assume the distance to be 11 kpc ( \citealt{FenderGarringtonMcKay1999,DhawanMirabelRodrguez2000,ZdziarskiRao2005}, but see also \citealt{ChapuisCorbel2004,KaiserGunnBrocksopp2004} for lower values)
\end{itemize}

\begin{figure*}
\resizebox{8.7cm}{!}{\includegraphics{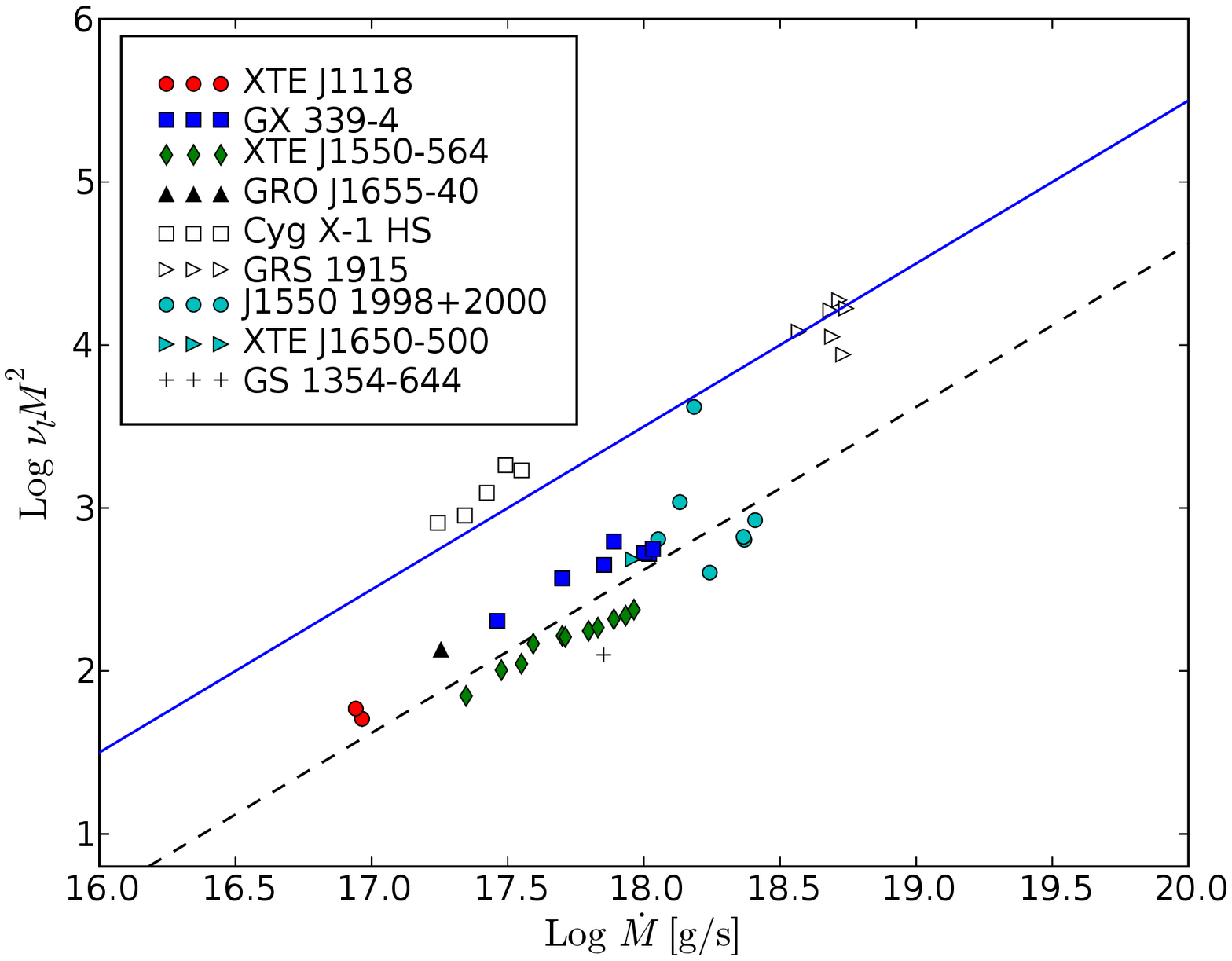}} \resizebox{8.7cm}{!}{\includegraphics{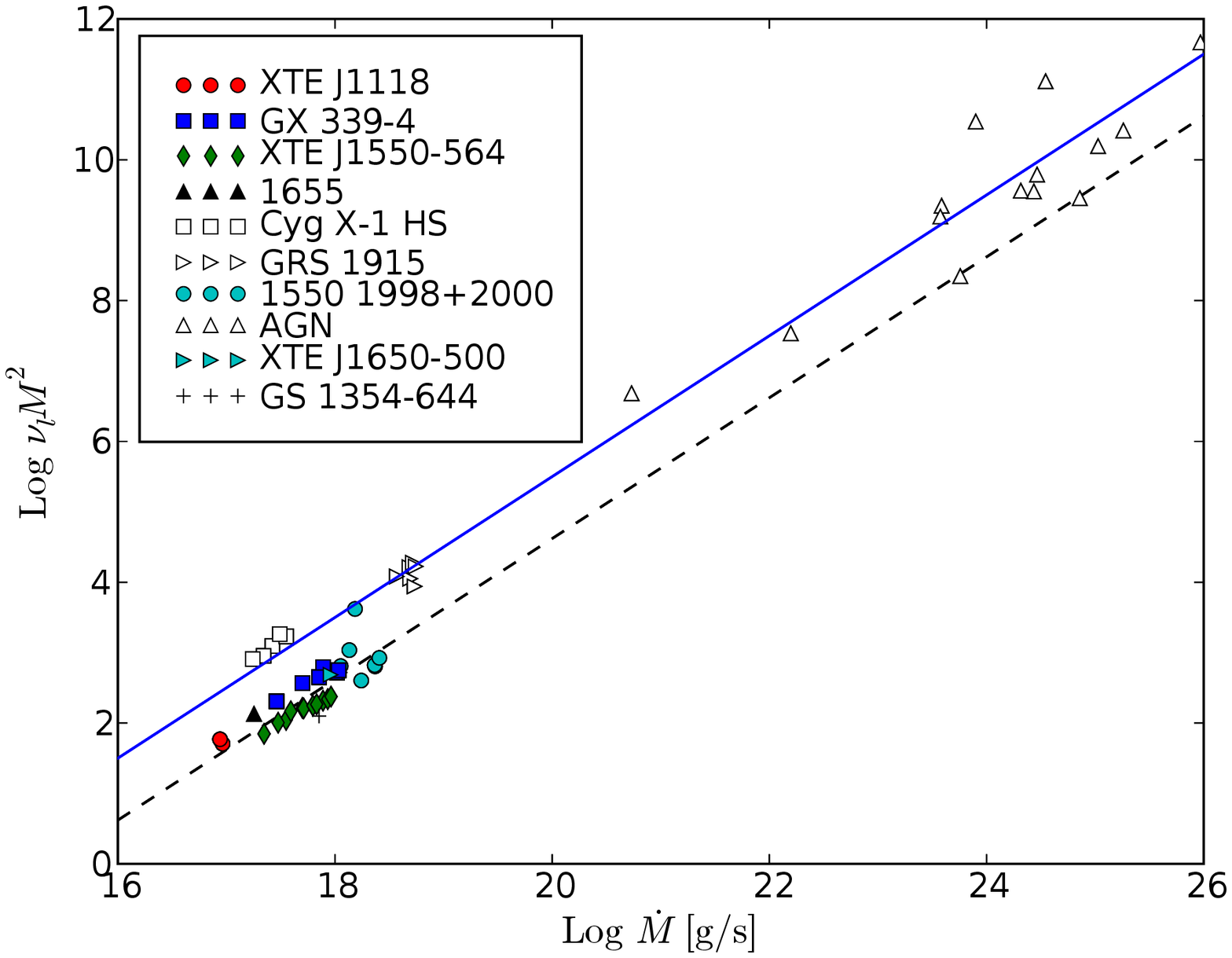}}
\caption{Left side: Our sample of stellar BHs. For Cyg X-1 we plot $\nu_{hb}$, while we show $\nu_l$ for the other objects. The lines indicate $\nu_l \propto \dot{M} M^{-2}$. While the upper line is a fit to soft-state objects  (including AGN, see \protect\cite{McHardyKoerding2006}), the lower is a fit to the hard-state XRBs only. Right side: Same as left side, but here we plot stellar and supermassive objects. In both panels soft-state and IMS objects are plotted with open symbols. \label{FigBHs}}
\end{figure*}

\subsection{Neutron Stars}
For all NSs we assume a mass of 1.4 $M_\odot$. 
\begin{itemize}
\item {\it low luminosity X-ray burster}: We include  1E 1724-304, GS 1826-24, SLX 1735-269 with $\nu_l$ measurements of  \citet{BelloniPsaltis2002}. The 2-10 keV X-ray flux has been obtained from archival RXTE PCA data. To obtain the bolometric luminosity and therefore the accretion rate, we use a bolometric correction of 2.5 found for atoll NSs \citep{MigliariFender2005b} also for these X-ray burster. 
We assume distances of 6.6 kpc for 1E 1724-304 \citep{BarbuyBicaOrtolani1998}, 5 kpc for GS 1826-24  \citep{ThompsonRothschildTomsick2005} and 8 kpc for SLX 1735-269 \citep{MolkovRevnivtsevLutovinov2005}.

\item {\it IGR J00291+5934}: The accreting millisecond X-ray pulsar IGR J00291+5934 went into outburst in 2004 \citep{ShawMowlaviRodriguez2005}. \citet{LinaresWijnands2006} have measured its frequency $\nu_l$, and we obtain 2-10 keV flux measurements from archival RXTE data. As V507 Cas is also in the field of view of the PCA, we assume that the quiescent flux can be attributed to the V507 Cas and subtract this from the measured flux of J00291+5934 during outburst. We assume a distance of 4 kpc \citep{GallowayMarkwardtMorgan2005}.

\item {\it atoll sources}: For the atoll sources we include 4U 1608-522 with $\nu_l$ measurements from \citet{vanStraaten2003} (distance of 3.4 kpc; \citealt{JonkerNelemans2004}). The 2-10 keV flux has been estimated from the 3-9 keV PCA count rate assuming a power law with $\Gamma = 1.6$. To obtain a bolometric luminosity we use a bolometric correction of 2.5. For 4U 1812-12 we use public RXTE data to obtain $\nu_l$ measurements and use the bolometric X-ray fluxes from \citet{BarretOliveOosterbroek2003}.  It should be noted that our measured values for $\nu_l$ are double of those of \citet{BarretOliveOosterbroek2003}. We assume a distance of 4 kpc \citep{CocchiBazzanoNatalucci2000}.

\item {\it Z-sources}: We include the Z-sources GX~340 (\citealt{JonkerWijnands1998}, distance of 9.5 kpc) and GX~5-1  (\citealt{WijnandsMendezPsaltis1998}, 7.4 kpc). The conversion factor from the 2-16 keV PCA count rate to bolometric luminosity of our Z-sources has been obtained from the Z-source GX~17+2 \citet{diSalvoStellaRobba2000}. The 0.1-200 keV flux given in the paper has been divided by the 2-16 keV PCA counts. For the Z-sources we use the lower kHz QPO as the frequency $\nu_l$ following \citet{BelloniPsaltis2002}.

\item {\it 4U~1728-34}: For 4U~1728-34 \citet{MigliariFender2005} measure the dependence of $\nu_h$ on the radio luminosity and find that it scales with $\nu_h \propto L_{Rad}^{0.7} \propto \dot{M}$. We include these points for comparison. We assume a distance of 4.6 kpc \citep{GallowayPsaltisChakrabarty2003}.
\end{itemize}

\subsection{AGN}
For AGN, we use the sample presented by \citet{UttleyMcHardy2005}. They give the high break frequencies, as well as the black hole masses and the bolometric luminosities. The majority of sources are Seyfert galaxies, and thus they are likely high- or very-high-state objects. We can therefore translate the bolometric luminosities directly to accretion rates.

\section{Results} \label{seresults}
For our AGN and the two stellar soft-state black holes we found in \citet{McHardyKoerding2006} that the measured high break frequencies $\nu_{hb}$ lie on a plane in the $\nu_{\mathrm hb}$, $\dot{M}$ and $M$ space:
\begin{equation}
\log \nu_{\mathrm hb} = \xi_{\mathrm acc} \log \dot{M} + \xi_{\mathrm m} \log M + b_{\nu}, \label{eqplane}
\end{equation}
where $\xi_{\mathrm acc}$ and  $\xi_{\mathrm m}$ denote the correlation indexes for the accretion rate and the black hole mass; $b_{\nu}$ denotes the constant offset. We found that $\xi_{\mathrm acc} = 0.98 \pm 0.15$ and  $\xi_{\mathrm m} =  -2.1 \pm 0.15$. Both parameters are within the uncertainties of integer value. We will therefore adopt the integer solution $(\xi_{\mathrm acc} = 1,\xi_{\mathrm m} =-2 )$ and will not present new fits here.  

The edge-on projection of the plane using our sample is shown in Fig.~\ref{FigBHs}. The two lines indicating $\nu_{l,hb} \propto \dot{M} M^{-2}$ ($\nu_{l,hb}$ denotes $\nu_l$ for hard-state objects and $\nu_{hb}$ for soft-state objects). The upper line is the fit to soft-state stellar BHs and supermassive BHs using integer valued parameters for the variability plane. For the adopted parameters and our units the constant offset $b_\nu$ for the soft-state objects is $b_\nu= -14.5 \pm 0.1$. This offset is different from the one given in \citet{McHardyKoerding2006}, as we use different units and fixed the parameters of the plane to integer values.

\begin{figure*}
\resizebox{8.7cm}{!}{\includegraphics{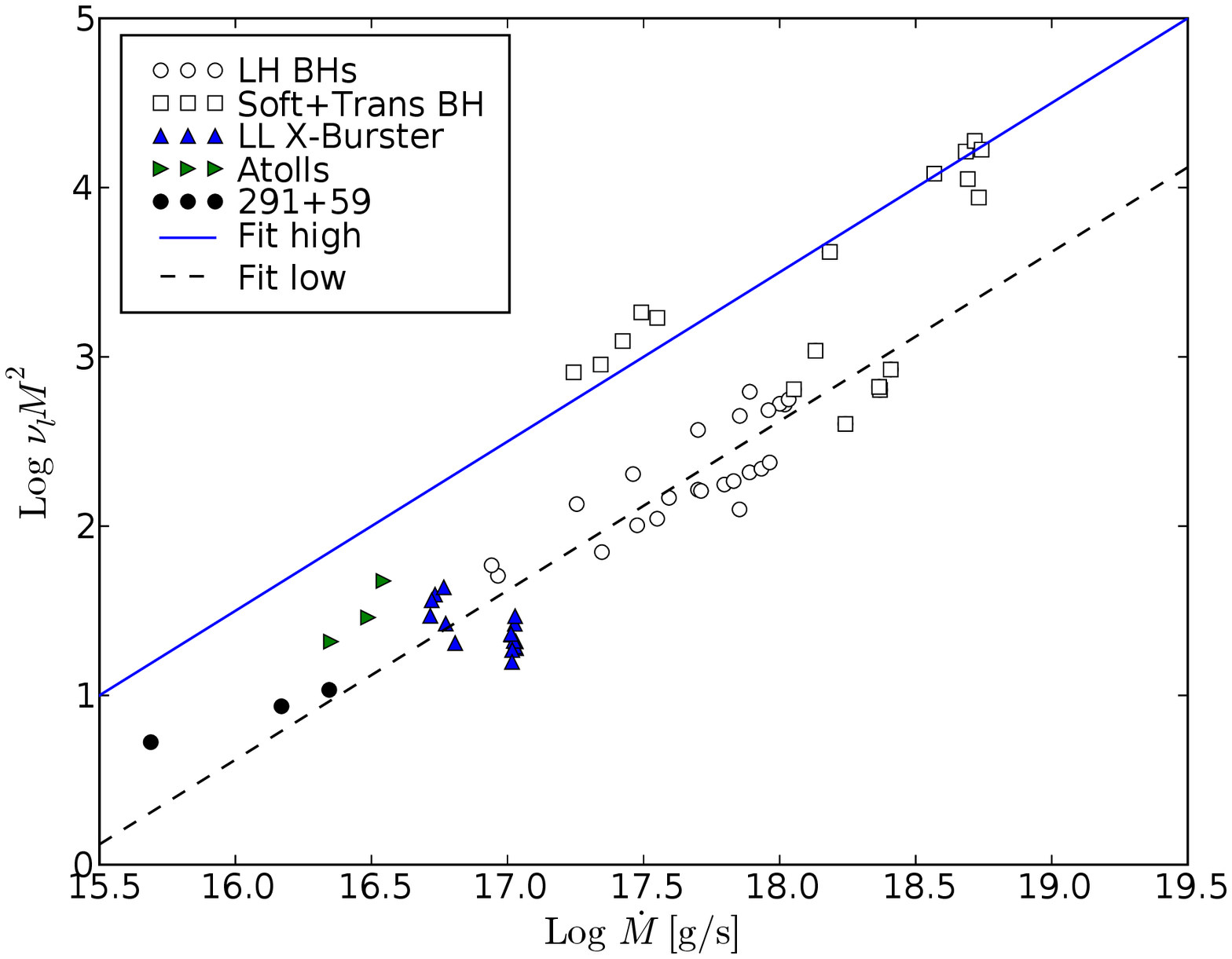}}
\resizebox{8.7cm}{!}{\includegraphics{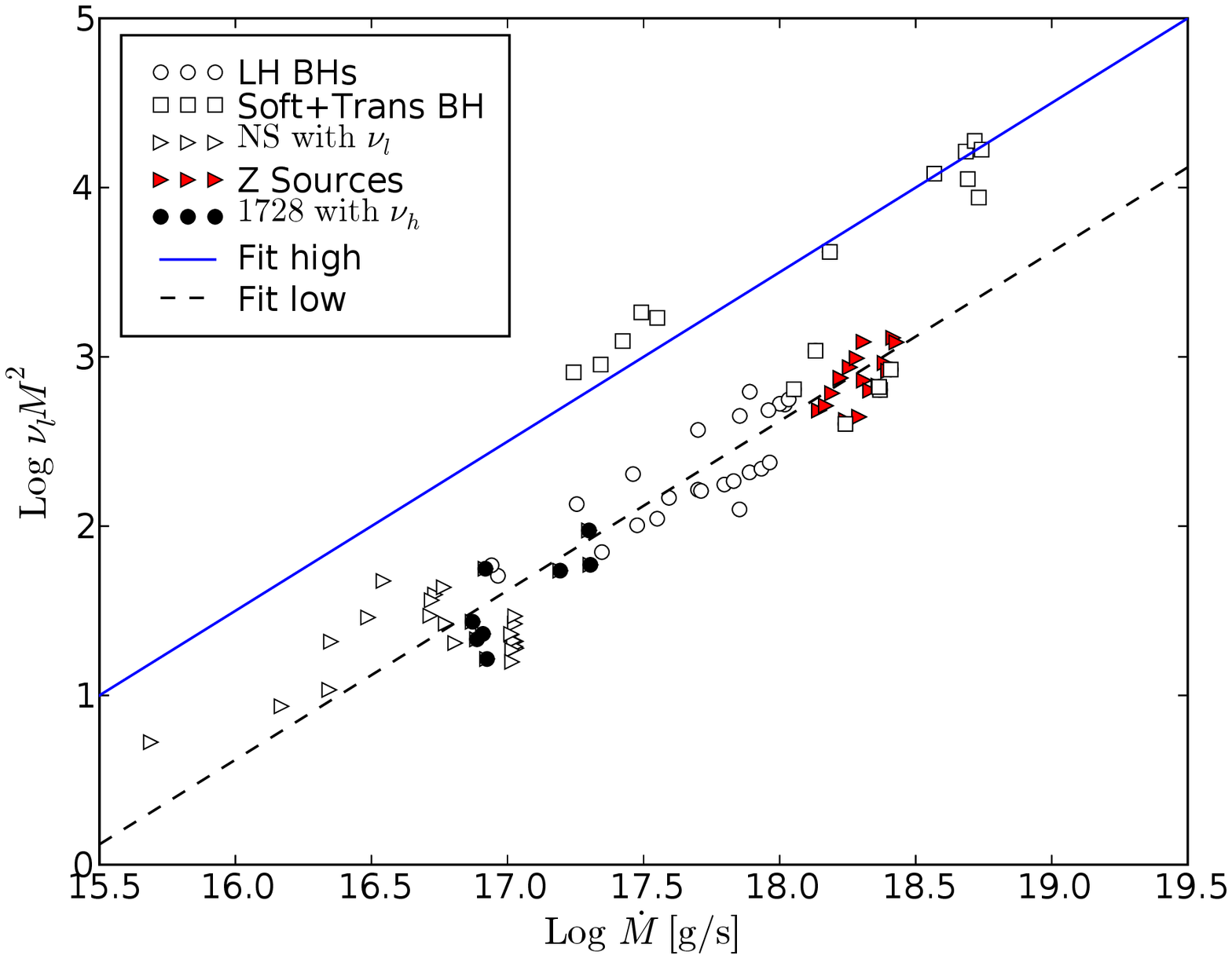}}
\caption{Left panel: NSs with measured $\nu_l$ in comparison with our stellar BH sample. The NS sample seems to follow the correlation, albeit with larger scatter. The upper line shows the correlation normalized to soft-state objects while the lower line is normalized for hard-state objects. Right panel: Inclusion of kHz QPOs of Z sources. While the kHz QPOs of Z sources are near the correlation this is not true for kHz QPOs of atoll sources. See text for a detailed discussion. }
\label{fiNSs}
\end{figure*}

\subsection{Stellar black holes}
On the left side of Fig.~\ref{FigBHs}, we show the projection of the variability plane "zoomed-in" on our sample of stellar BHs. For hard-state BHs we have argued that the frequency $\nu_l$ of the lower high-frequency Lorentzian corresponds to the high-frequency break $\nu_{hb}$ used for soft state. In Fig.~\ref{FigBHs} we also plot our hard-state BH sample. Similar to the scaling found in soft-state XRBs and AGN, we find that our hard-state BHs follow a scaling $\nu_l \propto \dot{M}$. However, the constant offset is different for hard and soft-state objects, and we also show a plane normalized to the hard-state objects only. The constant offset between the soft and the hard state scaling is $-0.9$ dex.

The uncertainties of the measured values shown in Fig.~\ref{FigBHs} are dominated by systematic errors (e.g., uncertainties in the accretion rate measure) as well as uncertainties of the black hole mass and distance of the black holes. Most uncertainties of all measurements of one source are coupled. Showing these coupled uncertainties as errorbars for the data-points is therefore misleading. The error budget consists of:
\begin{itemize}
\item Uncertainties of the primary parameters $\nu_l$ and the radio or X-ray flux: These uncertainties are different for each data-point. A typical value is $\leq 0.04$ dex.
\item Uncertainties in the BH mass and distance measurements: These uncertainties are shared for all data-points of a given source. Typical values are between 0.03 and 0.3 dex. This component dominates the total error budget, especially as they enter the variability plane quadratically and our two main sources (GX~339-4 and XTE J1550-564) have large uncertainties in the mass estimate (0.2 and 0.15 dex). 
\item Uncertainties of the accretion rate measure: This uncertainty has two components: Firstly, an uncertainty of the normalisation of the accretion rate measure, this would directly change the normalisation of all data-points. Secondly, uncertainties due to source peculiarities and the individual measurements. The exact value of these uncertainties is hard to access. In Sect.~\ref{seacc} we have argued that their combined effect is roughly $\sim 0.2$ dex.
\end{itemize}

The correlation of $\nu_l \propto \dot{M}$ is found in both hard-state BH that have several measurement (GX~339-4 and XTE J1550-564). The majority of the uncertainties just mentioned do not play a role if one only considers only a single source as they would only change the normalisation constant $b_\nu$. Thus, we can safely assume that there is a correlation between  $\nu_l$ and $\dot{M}$ for individual sources. Whether the constant offset $b_\nu$ is significantly different for hard and soft-state objects need to be verified. 

To access the value and uncertainty of the constant offset $b_\nu$ for our sample of sources we use two approaches: First we can compute the direct mean and sample variance of the sources around the correlation using the same weight for all data-points. We find  $b_\nu =  -15.39 \pm 0.04$ with a sample variance of $0.18$. However, as many of the errors of the data-points are coupled we might underestimate the uncertainty of the mean. 

To avoid the coupling of the uncertainties for different data-points of a given source we first treat every source individually and combine the different sources in a second step taking the estimated errors into account. To calculate the error on the mean $b_\nu$ we used the following steps:
\begin{enumerate}
\item Calculate $b_\nu$ for each source individually and measure the intrinsic scatter around the linear correlation through the sample variance $(\sigma_{int})$. 
\item Estimate the uncertainty of the measured $b_\nu$ for all sources. We set
\begin{equation}
\sigma_b^2 =  \sigma_{int}^2 + (2 \sigma_{D})^2 + (2 \sigma_{M})^2 + \sigma_{\dot{M}}^2,  
\end{equation}
where $\sigma_{D}$ is the uncertainty of the measurement of the distance, $\sigma_{M}$ the mass uncertainty and  $\sigma_{\dot{M}}$ the systematic uncertainty of the $\dot{M}$-measure.
\item Calculate that weighted mean and error of mean with the estimated uncertainties. Each source is weighted by their uncertainties and the number of measurements of the source.
\end{enumerate}
Using this method, we find $b_\nu =  -15.38 \pm 0.08$. The mean value of $b_\nu$ is in agreement with the value found using the sample variance.
Finally, we have to consider that the overall normalisation of the accretion rate measure has an uncertainty of $\sim 0.2$ dex. This uncertainty is shared by all data-points and all errors in the estimate of the normalisation effects the mean value of $b_\nu$ directly. If we include this uncertainty we obtain an overall error of $0.22$ dex. As the difference between the normalisation found for the hard and the soft state is 0.9 dex, we conclude that the difference between the hard and the soft-state scaling is significant.

Using this meathod we can also calculate the expected mean variance of the correlation given the assumed uncertainties. We find $\sigma = 0.38$, which is larger than the sample variance of $0.18$. The discrepance is mainly due to the coupled uncertainties, but we may also have overestimated the uncertainties of the accretion estimator or those of the distance and mass measurements. 

Only for GX~339-4 and XTR J1118+480 we have used the radio luminosity to obtain accretion rates while for the other hard-state objects we used the 2-10 keV flux. We have verified that Fig.~\ref{FigBHs} does not change significantly if we use the accretion rate measure based on the X-ray flux for all sources. This is not surprising, as the accretion rate measure used for the X-ray flux is partly based on the radio/X-ray correlation of GX~339-4. Thus, our findings are not affected by our choice of different the accretion rate estimators.

The outbursts of XTE~J1550-564 in 1998 and 2000 seem to follow a different track than the linear dependence seen in the hard-state outburst in 2002 and the other hard-state objects. The first two outbursts rise more steeply than the linear hard-state scaling. The outbursts seem to start at a frequency $\nu_l$ characteristic for a hard state and lead to values typically found for soft states. 
In these outbursts the photon index changes strongly during the observations. In 1998 the first observation in our sample has a photon index of $\Gamma = 1.53$ \citep{SobczakMcClintockRemillard2000} which changes continuesly to the last observation which has $\Gamma=1.98$, which is atypical for the hard state were one would expect the photon index to stay hard until the state transition. 
In the 2000 outburst the source is also softening from $\Gamma = 1.46$ to $1.7$ while the power law cutoff moves from 33 keV to 19 keV with a significant black-body component visible in the last observation used (a typical cut-off for the hard state is $\sim 80$ keV). Thus, it is likely that the source was already in an intermediate state in these outbursts.
This suggests that there is one scaling relation for the classical hard state and one with a higher normalisation for the soft state, but there is obviously another parameter that governs the transition between those scaling relations. During the transition from the hard to the soft state, the source increases its frequency fast and moves from the hard scaling to the soft one. 

For XRBs the transition from the soft to the hard scaling is probably best traced by the spectral index $\Gamma$. A strong correlation between timing properties and the photon index has been found in objects near the intermediate states, see e.g., \citet{Pottschmidt2002,VignarcaMigliariBelloni2003,KalemciTomsickBuxton2005}. A simple dependence on the photon index is however not directly applicable to AGN. The timing properties are also correlated to the hardness ratio see e.g., \citet{BelloniHomanCasella2005}. The hardness ratio can be generalised to the non-thermal fraction $f = \frac{L_{PL}}{L_{Disk}+L_{PL}}$, where $L_{Disk}$ is the luminosity in the multi-color black body component while $L_{PL}$ describes the luminosity of the power law component. This non-thermal fraction has similar properties as the hardness ratio but is also applicable to AGN (see \citealt{KoerdingJesterFender2006}).
Sources solidly in the soft state $(f=0)$ follow a track given by $\log \nu_{l,hb} = \log (\dot{M} M^{-2}) - 14.7$, while those in the hard state $(f=1)$ follow $\log \nu_{l,hb} = \log (\dot{M} M^{-2}) - 14.7 - 0.9$. In the transition between both states, the frequencies depend on the photon index and therefore on $f$. This suggests that the dependence of $\nu$ on $M$, $\dot{M}$ and $f$ can be approximated by:
\begin{equation}
\log \nu_{l,hb} = \log \dot{M} - 2 \log M - 14.7 - 0.9 \ \theta ( f )
\end{equation}
Where $\theta$ is a monotonic function with $\theta(0)=0$ and $\theta(1) = 1$. Unfortunately, with the readily available data of state transitions we are not able to fully constrain it. We note that this dependence may have to be further modified due to hysteresis effects similar to those seen in hardness-intensity diagrams.

\subsection{Neutron Stars}

In the left panel of Fig.~\ref{fiNSs} we show our sample of NSs with visible broad lower high-frequency Lorentzians $(L_l)$ together with the stellar BHs. The low luminosity X-ray burster and the accreting millisecond pulsar lie near the expected scaling for hard-state BHs -- but with increased scatter. Also our atoll sources are in a similar frequency range. However, while we could observe the linear dependency of $\nu_l$ on accretion rate in some single sources for BHs (ie., XTE J1550-564 and GX~339-4) this is harder for the NSs, since the accretion rate changes in our sample are small. The only object changing its accretion rate significantly is IGR J00291+5934. For that source, the slope seems to be slightly shallower than what is found for the BHs. However, as already mentioned, a second source (V507 Cas) is in the field of view of the PCA, so it is hard to measure correct fluxes for the pulsar. It may be that we are underestimating the lowest detected fluxes. As a sample the NS sources still scatter around the correlation found for hard-state BHs, albeit with a larger scatter.

\citet{PsaltisBelloni1999} present a tight correlation between the frequencies of the low frequency QPO $(\nu_{LF})$ and those of the lower kHz QPO in NSs and BHs. In the unified picture of \citet{BelloniPsaltis2002} the lower kHz QPO found in some NSs is identified with the broad lower high-frequency Lorentzian $L_l$ for BHs and weakly accreting NSs.
As the correlation between $\nu_l$ and $\dot{M} M^{-2}$ holds for some sources on the $\nu_{LF}$/$\nu_l$ correlation, it may be possible that one can extend our accretion rate $\nu_l$ correlation to all sources on the \citet{PsaltisBelloni1999} correlation.

We have just shown, that atoll sources with a measured frequency $\nu_l$ roughly follow the variability plane.  At higher accretion rates atoll sources show a lower kHz QPO at significantly higher frequencies than the observed values of $\nu_l$ at low accretion rates (when the lower kHz QPO is observed the broad lower high-frequency Lorentzian $L_l$ is not seen). However, the increase in the accretion rate is not sufficient to ensure that the lower kHz QPO is in agreement with the correlation.  This can be also shown by looking at $\nu_h$ for 1728-34. $\nu_h$ is the frequency corresponding to the broad "hump" Lorentzian ($L_h$) of the low frequency QPO. Thus $\nu_h$ and the frequency of the low frequency QPO are well correlated and have roughly similar values. As we show on the right side of Fig.~\ref{fiNSs}, $\nu_h$ is (at least for 1728) near the correlation. Thus, the lower kHz QPO will be above the correlation by a factor $\sim 15$. 

Z-sources have significantly higher accretion rates than atoll sources and often show kHz QPOs.  
The frequencies of the lower kHz QPOs of Z-sources are shown in Fig.~\ref{fiNSs}. The accretion rates are so large that the frequencies of the lower kHz QPO of Z-sources are near the hard-state correlation. However, Z-sources are very strongly accreting, e.g., they belong to a bright IMS. These sources should therefore follow the high state slope similar to the BH GRS~1915+105. The lower kHz QPO frequencies are so high, that they have the same order of magnitude as the Keplerian frequency at the surface of the neutron star $(\sim 1700 Hz)$. Thus, it may be that the frequencies start to saturate and are therefore slightly below the high state correlation. See also the discussion on WD and the corresponding Fig.~\ref{fiWDBreakup} where this effect may be even more pronounced.

The slope of the Z-sources alone seems to be steeper than the correlation found for BHs. It may be, that while the accretion rate does not change very much the sources makes a state transition from the analog of the hard IMS to the soft IMS (horizontal branch to normal branch). As we have seen for BHs, during state transitions the source moves rapidly form the hard-state scaling to the soft-state scaling. This rapid transition explain the steeper slope of Z-sources. 
Nevertheless, as the kHz QPOs of atoll sources do not fit on the correlation we can not be sure if it is not pure coincidence that the Z-sources seem to follow the correlation. For the correlation we therefore have to rely on sources with measured $\nu_l$.

We have mentioned in the introduction that the "parallel tracks" found in flux frequency diagrams may suggest that the frequency of the kHz QPO may be a better tracer of the instantaneous accretion rate than the X-ray luminosity \citep{Klis2001}. We are currently unable to use any frequency as a direct tracer of the accretion rate as the conversion factors are yet unknown. As we have shown it is likely that the frequency ($\nu_l$) does not only depend on the accretion rate but also on the state of the source. Thus, at least $\nu_l$ cannot be used as a measure of the accretion rate for all times if one does not have detailed information on the state of the source. If the frequency of the lower high-frequency Lorentzian does not only depend on the accretion rate but also on an additional parameter describing the accretion state, one could explain the parallel tracks. On short timescales the main parameter is either the accretion rate or the accretion rate and the accretion state parameter are strongly coupled creating the well correlated tracks. On longer timescales the relationship of the two parameter can change so that one observes parallel tracks. 

\begin{figure}
\resizebox{8.7cm}{!}{\includegraphics{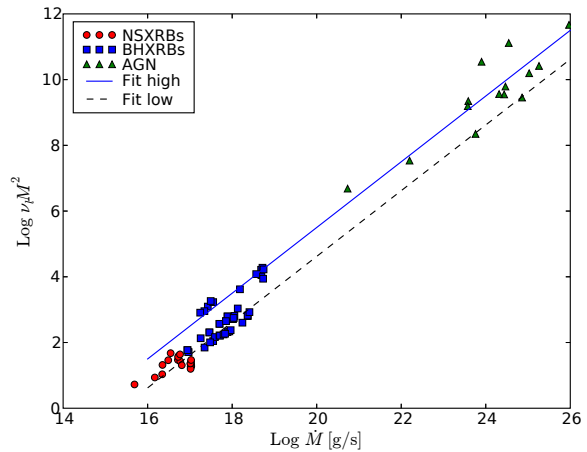}}
\caption{Edge-on projection of the variability plane including AGN, stellar BHs and NSs with directly measured $\nu_l$.}
\label{fiNSBH}
\end{figure}

The final edge-on projection of the variability plane including BH XRBs, NS XRBs with directly measured $\nu_l$ and AGN is shown in Fig.~\ref{fiNSBH}. As a sample the NSs extend the correlation toward lower accretion rates.

\subsection{Comments on DNOs for accreting white dwarfs}
\begin{figure}
\resizebox{8.7cm}{!}{\includegraphics{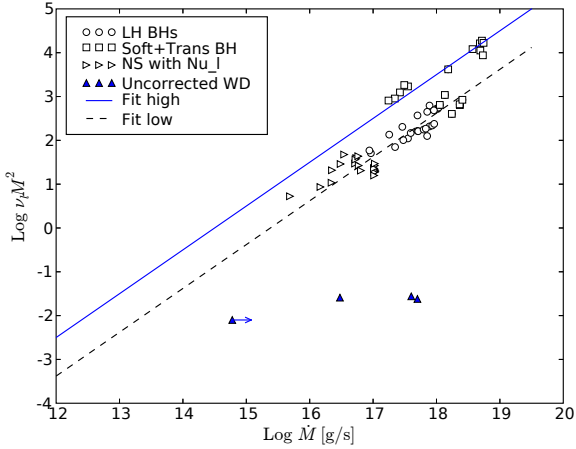}}
\caption{Edge-on projection of the variability plane compared to our sample of WDs. The uncertainty on the accretion rates in a factor 2-3, so in logspace 0.3-0.5 dex.}
\label{fiWD}
\end{figure}

\begin{table}
\caption{Parameters of our WD sample.}
\begin{center}
\begin{tabular}{lllll}
\hline
 & $\dot{M}$ [g/s] & $\nu_{\mathrm DNO}$ Hz& $\log \nu M^2$ & $(\log \nu M^2)_{\mathrm pred}$   \\ 
\hline \hline
IX Vel & $5 \times 10^{17}$ & 3.7 $ \times 10^{-2}$  & -1.62 & 3 \\ 
UX UMa & $4 \times 10^{17}$ & 3.4  $\times 10^{-2}$ & -1.56 & 2.9 \\
OY Car & $3\times 10^{16}$ & 5.5 $\times 10^{-2}$  & -1.59 & 1.8 \\
VW Hyi & $ \gtrsim 6 \times 10^{14}$ & $\sim 1 \times 10^{-2}$ & -2.1 & 0.1 \\
\hline
\end{tabular}
\end{center}
Masses: IX Vel: 0.8 $M_\odot$, UX UMa: 0.9 $M_\odot$, OY Car: 0.69 $M_\odot$ and VW Hyi: 0.84 $M_\odot$.
Refs: IX Vel: \citet{BeuermannThomas1990}, UX UMa: \citet{FroningLongKnigge2003,SuleimanovNeustroevBorisov2004,BaptistaHorneWade1998,KniggeDrakeLong1998}, OY Car: \citet{MarshHorne1998,PrattHassallNaylor1999,WoodHorneBerriman1989}, VW Hyi: \citet{WarnerWoudt2006,PandelHowell2003,SionChengSparks1997}
\label{tabCVs}
\end{table}

The aforementioned correlation of the frequencies of the LF QPOs with those of the lower kHz QPO \citep{PsaltisBelloni1999} has also been extended to include CVs \citep[e.g.,][]{WarnerWoudtPretorius2003}. Here, the dwarf-nova oscillation (DNO) frequency corresponds to the frequency $\nu_l$ of the lower high-frequency Lorentzian. In table~\ref{tabCVs} we tabulate the parameters of four well known systems. 
The accretion rates of CVs have higher uncertainties than those of NSs or BHs, as the majority of the bolometric luminosity is not emitted in the well accessable X-ray band. We assume that the accretion rates are only correct up to a factor of 2-3. Especially the accretion rate for VW Hyi is uncertain, as we use the quiescent value as a lower limit to the accretion rate, while the DNO frequencies are measured during the end of the decline from an outburst. Thus, the accretion rate will be approximately the quiescent rate.  Nevertheless, observed values of $\log \nu M^2$ are mostly around -1.6, while we expect values between 0.1 and 3 (assuming that they follow the soft-state scaling). Thus, in the current form, the correlation is not valid for CVs (see Fig.~\ref{fiWD}). We note that, for CVs, the DNO frequency rises with accretion rate \citep{WoudtWarner2002}, similar to the expected behaviour of the correlation.

\begin{figure}
\resizebox{8.7cm}{!}{\includegraphics{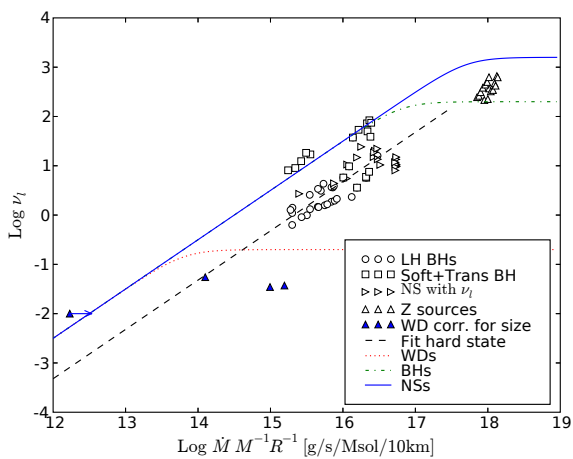}}
\caption{Projection of the variability plane onto the $\nu_l$ $\frac{\dot{M}}{M R}$ plane. Besides the scaling relation expected for hard-state objects we also show the breakup frequency of WDs ($\sim$ 5 Hz) and NSs ($\sim $1700 Hz) and the frequency at the innermost stable orbit of BHs ($\sim 240$ Hz). Sources can only follow the correlation until the frequencies are comparable to the breakup frequency, there the correlation need to level off to the constant breakup frequency.}
\label{fiWDBreakup}
\end{figure}

However, up to now we have not considered that a white dwarf is significantly larger than a neutron star or a black hole.
It is not evident how to correct the correlation for the larger radius. For BHs the radius is linearly related to the BH mass. Thus, the correlation found for BHs suggests that one or both mass terms of the correlation correspond to a radius. If one of the mass terms corresponds to a radius, the correlation reads $\nu_l \propto \frac{\dot{M}}{M R}$, where $R$ is the radius of the central object. This correlation can be read as a correlation between $\nu_l$ and the power liberated in the accretion disk around the central object divided by $M^2$. The power liberated in a standard accretion disk around a central object with radius $R$ is $R/(6 G M/c^2)$ times smaller than the power liberated in a disk around a non-rotating BH, the one radius term corrects for this fact. A white dwarf of average mass has a radius roughly $\sim 500$ times larger than that of a NS. This would move the expected position of OY Car and VW Hyi near our correlations (see Fig.~\ref{fiWDBreakup}). Their positions are in agreement with the correlation given the large uncertainties on the accretion rates. IX Vel and UX UMa, however, are still far from the correlation. Those two sources are nova-likes, with high accretion rates. If the correlation would hold for all those sources, their frequencies would exceed the breakup frequency of the white dwarf $(\sim 5)$ Hz. If the DNO frequency (or $\nu_l$) is somehow related to the Keplerian rotation this is impossible. Thus, it is likely that the sources follow the correlation until the frequencies are comparable to the fastest possible frequency (breakup frequency for WDs and the frequency of the innermost stable orbit for BHs). For these high accretion rates the frequencies can not increase any further and stay constant around the fastest possible frequency. Schematic frequency tracks for WDs, BHs and NSs are shown in Fig.~\ref{fiWDBreakup}. The shown lines should only illustrate possible frequency tracks. As the break up frequencies are reached at high luminosities the shown tracks are normalised to the high state correlation. The shown function is a smooth broken power law which is linear with $\dot{M}$ at low accretion rates and constant at high accretion rates ($\nu \propto \dot{M} (1+(\dot{M}/\dot{M}_0)^2)^{-0.5}$). If one assumes that both mass terms of the correlation correspond to radii, none of the measured WDs fall on the expected correlation. Thus, we consider it to be more likely that only one mass term corresponds to a radius. In summary, while dwarf novae systems seem to lie near the correlation if one takes the size of the central object into account, nova-like objects do not seem to follow the correlation as their frequencies would exceed the break up frequency of the WD.

\section{Discussion}
In the previous section we found that the variability plane found in soft-state accreting black holes \citep{McHardyKoerding2006} can be extended to hard-state stellar black holes (see Fig.~\ref{FigBHs}). The frequencies depend on the accretion rate $(\dot{M})$ and the black hole mass $(M)$ as:
\begin{equation}
\left(\frac{\nu_{\mathrm l}}{\mbox{Hz}}\right) \left(\frac{M}{10 M_\odot} \right) \approx 2.2 \left(\frac{\dot{M}}{0.01 \dot{M}_{\mathrm Edd}}\right), \label{eqfinal}
\end{equation}
for soft-state objects; for hard-state objects the proportionality factor is $\sim 0.3$. It is even possible to extend the correlation to NS with directly measured frequencies $\nu_l$ (see Fig.~\ref{fiNSBH}). Here, we discuss the physical implications of these findings.

This dependence on the accretion rate as well as on the black hole mass can be simplified by assuming that the accreting black hole is scale invariant (e.g., \citealt{MerloniHeinzdiMatteo2003,FalckeKoerdingMarkoff2004}).  
If the frequency arises through Keplerian motion or any other process scaling with the size of the central object, the left side of the equation is invariant of black hole mass. The right side denotes a simple linear dependence on the accretion rate in Eddington units. 

While a linear dependency on the accretion rate is seen if a source is in the canonical hard or soft state this is not the case for sources in intermediate states or near the state transition. For example, we have seen that XTE~J1550-564 follows the linear scaling in its hard outburst in 2002, while the frequencies rise faster than linear for the other outbursts. Thus, besides accretion rate there must be at least a second parameter governing the observed frequencies ($\nu_l,\nu_{hb}$). Similar to the hysteresis found for state changes, (see e.g, the hardness intensity diagrams of XRB outbursts,) there seems to be a hysteresis effect for the frequencies (see Fig.~7 in \citealt{BelloniHomanCasella2005}).

In case the high-frequency break originates from Keplerian motion or a frequency that is related to the Keplerian frequency, we can use eq.~(\ref{eqfinal}) to obtain the dependence of the active radius $r_a$, which corresponds to the break in the PSD, on the accretion rate. This yields: 
\begin{equation}
\frac{r_a}{R_G} = 625 \left(\frac{\dot{M}}{0.01 \dot{M}_{Edd}}\right)^{-2/3},
\end{equation}
where $R_G$ is the gravitational radius $R_G = G M /c^2$.
The correlation with the accretion rate therefore implies that the radius associated with the break moves inwards with $\dot{M}^{-2/3}$. Interestingly, such a behaviour may be observed in GRS 1915+105. \cite{BelloniMendezKing1997} reports that the disk radius associated with the region emitting the measured X-ray spectrum changes from 70 km to roughly 20 km while the RXTE/PCA count rate increases by a factor of $\sim 5$. If the bolometric correction does not change significantly between the different observations, this suggests that the emitting radius moves inwards as $r^{-0.75}$ compared to our prediction of $r^{-0.67}$.

It is tempting to associate the active region creating the high-frequency break with the transition from the geometrically thin standard disk to the optically thin inefficient accretion flow. However, most of our objects are soft-state objects. Within the soft state it is usually assumed that the optically thick, geometrically thin accretion disk reaches up to the last stable orbit. The luminosities of a given source in the soft state can change by a factor 15 (2-30\% Eddington), so the inner disk radius would have to change by a factor 6. This is not in agreement with the redii deduced from X-ray spectral fitting nor with observed scaling of the X-ray luminosities with disc temperature as $T^4$ \citep[e.g.,][]{GierlinskiDone2004}. Furthermore, the observed frequencies are far too low to originate from a Keplerian motion at the inner edge of the accretion disk (innermost stable orbit). We further note that for a normal standard disk, viscous timescales are too long ($\sim 2$ s) to explain the high frequencies and seem to have a very low dependency on accretion rate ($\alpha^{-4/5} M^{-3/10}$). It may well be that the break frequency is indeed associated with a Keplerian motion. However, as we have argued, the corresponding radius does not seem to be directly associated with the inner edge of the standard disk. Maybe this radius only reaches the innermost radii for highly accreting objects like Z-sources with their lower kHz QPO.

We have seen that NSs with measured $\nu_l$ seem to follow the suggested correlation. At higher accretion rate the Z-sources have lower frequencies than expected. This discrepancy may arise from the slightly larger size of the NS with its boundary layer compared to a BH with similar mass. While increasing the accretion rate, the emitting region cannot move inwards as expected as it is stopped by the boundary layer. This may also be the reason for the relatively low DNO frequencies of strongly accreting white dwarfs (see Fig.~\ref{fiWDBreakup}).

\section{Conclusion}
We have shown that the variability plane found in AGN and soft-state XRBs ($\nu_{\mathrm hb} M \propto \dot{M}/\dot{M}_{\mathrm Edd}$) can be extended to hard-state BH XRBs and weakly accreting NSs. For hard-state objects the frequency of the lower high-frequency Lorentzian $(\nu_l)$ corresponds to the high-frequency break in AGN and soft-state XRBs. For white dwarfs, the situation seems to be more complicated as we need to correct of the WD radius. While some dwarf novae seem to lie near the correlation (within the large uncertainties), strongly accreting nova-like systems do not seem to follow the correlation as the frequencies would otherwise exceed the breakup frequency of the WD. In case that the high-break frequency in the PSD is related to a Keplerian orbit around the black hole, we find that this radius in gravitational radii depends on the accretion rate as $(\dot{M}/\dot{M}_{\mathrm Edd})^{(-2/3)}$. However, the orbit corresponding to the high break is unlikely directly related to the radius where the standard optically thick disk turns into an optically thin inefficient accretion flow. 

\section*{Acknowledgements}
We thank Tom Maccarone for helpful discussions. EGK would like to thank Joern Wilms for discussions on a related topic. We are grateful for constructive comments by our referee.
EGK acknowledges funding via a Marie Curie Intra-European Fellowship under
contract no. MEIF-CT-2006-024668.

\label{lastpage}

\bibliographystyle{mn2e}

\begin{thebibliography}{}

\bibitem[\protect\citeauthoryear{{Axelsson}, {Borgonovo} \&
  {Larsson}}{{Axelsson} et~al.}{2006}]{AxelssonBorgonovoLarsson2006}
{Axelsson} M.,  {Borgonovo} L.,    {Larsson} S.,  2006, \aap, 452, 975

\bibitem[\protect\citeauthoryear{{Baptista}, {Horne}, {Wade}, {Hubeny}, {Long}
  \& {Rutten}}{{Baptista} et~al.}{1998}]{BaptistaHorneWade1998}
{Baptista} R.,  {Horne} K.,  {Wade} R.~A.,  {Hubeny} I.,  {Long} K.~S.,
  {Rutten} R.~G.~M.,  1998, \mnras, 298, 1079

\bibitem[\protect\citeauthoryear{{Barbuy}, {Bica} \& {Ortolani}}{{Barbuy}
  et~al.}{1998}]{BarbuyBicaOrtolani1998}
{Barbuy} B.,  {Bica} E.,    {Ortolani} S.,  1998, \aap, 333, 117

\bibitem[\protect\citeauthoryear{{Barret}, {Olive} \& {Oosterbroek}}{{Barret}
  et~al.}{2003}]{BarretOliveOosterbroek2003}
{Barret} D.,  {Olive} J.~F.,    {Oosterbroek} T.,  2003, \aap, 400, 643

\bibitem[\protect\citeauthoryear{{Belloni}, {Colombo}, {Homan}, {Campana} \&
  {van der Klis}}{{Belloni} et~al.}{2002a}]{BelloniColomboHoman2002}
{Belloni} T.,  {Colombo} A.~P.,  {Homan} J.,  {Campana} S.,    {van der Klis}
  M.,  2002a, \aap, 390, 199

\bibitem[\protect\citeauthoryear{{Belloni}, {Homan}, {Casella}, {van der Klis},
  {Nespoli}, {Lewin}, {Miller} \& {M{\'e}ndez}}{{Belloni}
  et~al.}{2005}]{BelloniHomanCasella2005}
{Belloni} T.,  {Homan} J.,  {Casella} P.,  {van der Klis} M.,  {Nespoli} E.,
  {Lewin} W.~H.~G.,  {Miller} J.~M.,    {M{\'e}ndez} M.,  2005, \aap, 440, 207

\bibitem[\protect\citeauthoryear{{Belloni}, {Mendez}, {King}, {van der Klis} \&
  {van Paradijs}}{{Belloni} et~al.}{1997}]{BelloniMendezKing1997}
{Belloni} T.,  {Mendez} M.,  {King} A.~R.,  {van der Klis} M.,    {van
  Paradijs} J.,  1997, \apjl, 488, L109

\bibitem[\protect\citeauthoryear{{Belloni}, {Psaltis} \& {van der
  Klis}}{{Belloni} et~al.}{2002b}]{BelloniPsaltis2002}
{Belloni} T.,  {Psaltis} D.,    {van der Klis} M.,  2002b, \apj, 572, 392

\bibitem[\protect\citeauthoryear{{Beuermann} \& {Thomas}}{{Beuermann} \&
  {Thomas}}{1990}]{BeuermannThomas1990}
{Beuermann} K.,  {Thomas} H.-C.,  1990, \aap, 230, 326

\bibitem[\protect\citeauthoryear{{Casares}, {Zurita}, {Shahbaz}, {Charles} \&
  {Fender}}{{Casares} et~al.}{2004}]{CasaresZuritaShahbaz2004}
{Casares} J.,  {Zurita} C.,  {Shahbaz} T.,  {Charles} P.~A.,    {Fender} R.~P.,
   2004, \apjl, 613, L133

\bibitem[\protect\citeauthoryear{{Chapuis} \& {Corbel}}{{Chapuis} \&
  {Corbel}}{2004}]{ChapuisCorbel2004}
{Chapuis} C.,  {Corbel} S.,  2004, \aap, 414, 659

\bibitem[\protect\citeauthoryear{{Chaty}, {Haswell}, {Malzac}, {Hynes},
  {Shrader} \& {Cui}}{{Chaty} et~al.}{2003}]{ChatyHaswellMalzac2003}
{Chaty} S.,  {Haswell} C.~A.,  {Malzac} J.,  {Hynes} R.~I.,  {Shrader} C.~R.,
   {Cui} W.,  2003, \mnras, 346, 689

\bibitem[\protect\citeauthoryear{{Cocchi}, {Bazzano}, {Natalucci}, {Ubertini},
  {Heise}, {Kuulkers} \& {in't Zand}}{{Cocchi}
  et~al.}{2000}]{CocchiBazzanoNatalucci2000}
{Cocchi} M.,  {Bazzano} A.,  {Natalucci} L.,  {Ubertini} P.,  {Heise} J.,
  {Kuulkers} E.,    {in't Zand} J.~J.~M.,  2000, in {McConnell} M.~L.,  {Ryan}
  J.~M.,  eds, American Institute of Physics Conference Series {Bright X-Ray
  Bursts from 1E1724-3045 in Terzan 2}.
pp 217--

\bibitem[\protect\citeauthoryear{{Cui}, {Zhang}, {Focke} \& {Swank}}{{Cui}
  et~al.}{1997}]{CuiZhangFocke1997}
{Cui} W.,  {Zhang} S.~N.,  {Focke} W.,    {Swank} J.~H.,  1997, \apj, 484, 383

\bibitem[\protect\citeauthoryear{{Dhawan}, {Mirabel} \& {Rodr{\'
  i}guez}}{{Dhawan} et~al.}{2000}]{DhawanMirabelRodrguez2000}
{Dhawan} V.,  {Mirabel} I.~F.,    {Rodr{\' i}guez} L.~F.,  2000, \apj, 543, 373

\bibitem[\protect\citeauthoryear{{Di Salvo}, {Stella}, {Robba}, {van der Klis},
  {Burderi}, {Israel}, {Homan}, {Campana}, {Frontera} \& {Parmar}}{{Di Salvo}
  et~al.}{2000}]{diSalvoStellaRobba2000}
{Di Salvo} T.,  {Stella} L.,  {Robba} N.~R.,  {van der Klis} M.,  {Burderi} L.,
   {Israel} G.~L.,  {Homan} J.,  {Campana} S.,  {Frontera} F.,    {Parmar}
  A.~N.,  2000, \apjl, 544, L119

\bibitem[\protect\citeauthoryear{{Edelson} \& {Nandra}}{{Edelson} \&
  {Nandra}}{1999}]{EdelsonNandra1999}
{Edelson} R.,  {Nandra} K.,  1999, \apj, 514, 682

\bibitem[\protect\citeauthoryear{{Esin}, {McClintock} \& {Narayan}}{{Esin}
  et~al.}{1997}]{EsinMcClintockNarayan1997}
{Esin} A.~A.,  {McClintock} J.~E.,    {Narayan} R.,  1997, \apj, 489, 865+

\bibitem[\protect\citeauthoryear{{Falcke}, {K\"ording} \& {Markoff}}{{Falcke}
  et~al.}{2004}]{FalckeKoerdingMarkoff2004}
{Falcke} H.,  {K\"ording} E.,    {Markoff} S.,  2004, \aap, 414, 895

\bibitem[\protect\citeauthoryear{{Fender} \& {Belloni}}{{Fender} \&
  {Belloni}}{2004}]{FenderBelloni2004}
{Fender} R.,  {Belloni} T.,  2004, \araa, 42, 317

\bibitem[\protect\citeauthoryear{{Fender}, {Garrington}, {McKay}, {Muxlow},
  {Pooley}, {Spencer}, {Stirling} \& {Waltman}}{{Fender}
  et~al.}{1999}]{FenderGarringtonMcKay1999}
{Fender} R.~P.,  {Garrington} S.~T.,  {McKay} D.~J.,  {Muxlow} T.~W.~B.,
  {Pooley} G.~G.,  {Spencer} R.~E.,  {Stirling} A.~M.,    {Waltman} E.~B.,
  1999, \mnras, 304, 865

\bibitem[\protect\citeauthoryear{{Foellmi}, {Depagne}, {Dall} \&
  {Mirabel}}{{Foellmi} et~al.}{2006}]{FoellmiDepagneDall2006}
{Foellmi} C.,  {Depagne} E.,  {Dall} T.~H.,    {Mirabel} I.~F.,  2006, \aap,
  457, 249

\bibitem[\protect\citeauthoryear{{Ford}, {van der Klis}, {M{\'e}ndez},
  {Wijnands}, {Homan}, {Jonker} \& {van Paradijs}}{{Ford}
  et~al.}{2000}]{Ford2000}
{Ford} E.~C.,  {van der Klis} M.,  {M{\'e}ndez} M.,  {Wijnands} R.,  {Homan}
  J.,  {Jonker} P.~G.,    {van Paradijs} J.,  2000, \apj, 537, 368

\bibitem[\protect\citeauthoryear{{Froning}, {Long} \& {Knigge}}{{Froning}
  et~al.}{2003}]{FroningLongKnigge2003}
{Froning} C.~S.,  {Long} K.~S.,    {Knigge} C.,  2003, \apj, 584, 433

\bibitem[\protect\citeauthoryear{{Galloway}, {Markwardt}, {Morgan},
  {Chakrabarty} \& {Strohmayer}}{{Galloway}
  et~al.}{2005}]{GallowayMarkwardtMorgan2005}
{Galloway} D.~K.,  {Markwardt} C.~B.,  {Morgan} E.~H.,  {Chakrabarty} D.,
  {Strohmayer} T.~E.,  2005, \apjl, 622, L45

\bibitem[\protect\citeauthoryear{{Galloway}, {Psaltis}, {Chakrabarty} \&
  {Muno}}{{Galloway} et~al.}{2003}]{GallowayPsaltisChakrabarty2003}
{Galloway} D.~K.,  {Psaltis} D.,  {Chakrabarty} D.,    {Muno} M.~P.,  2003,
  \apj, 590, 999

\bibitem[\protect\citeauthoryear{{Gierli{\'n}ski} \& {Done}}{{Gierli{\'n}ski}
  \& {Done}}{2004}]{GierlinskiDone2004}
{Gierli{\'n}ski} M.,  {Done} C.,  2004, \mnras, 347, 885

\bibitem[\protect\citeauthoryear{{Green}, {McHardy} \& {Lehto}}{{Green}
  et~al.}{1993}]{GreenMcHardyLehto1993}
{Green} A.~R.,  {McHardy} I.~M.,    {Lehto} H.~J.,  1993, \mnras, 265, 664

\bibitem[\protect\citeauthoryear{{Greene}, {Bailyn} \& {Orosz}}{{Greene}
  et~al.}{2001}]{GreeneBailynOrosz2001}
{Greene} J.,  {Bailyn} C.~D.,    {Orosz} J.~A.,  2001, \apj, 554, 1290

\bibitem[\protect\citeauthoryear{{Hjellming} \& {Rupen}}{{Hjellming} \&
  {Rupen}}{1995}]{HjellmingRupen1995}
{Hjellming} R.~M.,  {Rupen} M.~P.,  1995, \nat, 375, 464

\bibitem[\protect\citeauthoryear{{Homan} \& {Belloni}}{{Homan} \&
  {Belloni}}{2005}]{HomanBelloni2005}
{Homan} J.,  {Belloni} T.,  2005, \apss, 300, 107

\bibitem[\protect\citeauthoryear{{Homan}, {Klein-Wolt}, {Rossi}, {Miller},
  {Wijnands}, {Belloni}, {van der Klis} \& {Lewin}}{{Homan}
  et~al.}{2003}]{HomanKlein-WoltRossi2003}
{Homan} J.,  {Klein-Wolt} M.,  {Rossi} S.,  {Miller} J.~M.,  {Wijnands} R.,
  {Belloni} T.,  {van der Klis} M.,    {Lewin} W.~H.~G.,  2003, \apj, 586, 1262

\bibitem[\protect\citeauthoryear{{Hynes}, {Steeghs}, {Casares}, {Charles} \&
  {O'Brien}}{{Hynes} et~al.}{2003}]{HynesSteeghsCasares2003}
{Hynes} R.~I.,  {Steeghs} D.,  {Casares} J.,  {Charles} P.~A.,    {O'Brien} K.,
   2003, \apjl, 583, L95

\bibitem[\protect\citeauthoryear{{Hynes}, {Steeghs}, {Casares}, {Charles} \&
  {O'Brien}}{{Hynes} et~al.}{2004}]{HynesSteeghsCasares2004}
{Hynes} R.~I.,  {Steeghs} D.,  {Casares} J.,  {Charles} P.~A.,    {O'Brien} K.,
   2004, \apj, 609, 317

\bibitem[\protect\citeauthoryear{{Jonker} \& {Nelemans}}{{Jonker} \&
  {Nelemans}}{2004}]{JonkerNelemans2004}
{Jonker} P.~G.,  {Nelemans} G.,  2004, \mnras, 354, 355

\bibitem[\protect\citeauthoryear{{Jonker}, {Wijnands}, {van der Klis},
  {Psaltis}, {Kuulkers} \& {Lamb}}{{Jonker} et~al.}{1998}]{JonkerWijnands1998}
{Jonker} P.~G.,  {Wijnands} R.,  {van der Klis} M.,  {Psaltis} D.,  {Kuulkers}
  E.,    {Lamb} F.~K.,  1998, \apjl, 499, L191

\bibitem[\protect\citeauthoryear{{Kaiser}, {Gunn}, {Brocksopp} \&
  {Sokoloski}}{{Kaiser} et~al.}{2004}]{KaiserGunnBrocksopp2004}
{Kaiser} C.~R.,  {Gunn} K.~F.,  {Brocksopp} C.,    {Sokoloski} J.~L.,  2004,
  \apj, 612, 332

\bibitem[\protect\citeauthoryear{{Kalemci}, {Tomsick}, {Buxton}, {Rothschild},
  {Pottschmidt}, {Corbel}, {Brocksopp} \& {Kaaret}}{{Kalemci}
  et~al.}{2005}]{KalemciTomsickBuxton2005}
{Kalemci} E.,  {Tomsick} J.~A.,  {Buxton} M.~M.,  {Rothschild} R.~E.,
  {Pottschmidt} K.,  {Corbel} S.,  {Brocksopp} C.,    {Kaaret} P.,  2005, \apj,
  622, 508

\bibitem[\protect\citeauthoryear{{Kitamoto}, {Tsunemi}, {Pedersen}, {Ilovaisky}
  \& {van der Klis}}{{Kitamoto} et~al.}{1990}]{KitamotoTsunemiPedersen1990}
{Kitamoto} S.,  {Tsunemi} H.,  {Pedersen} H.,  {Ilovaisky} S.~A.,    {van der
  Klis} M.,  1990, \apj, 361, 590

\bibitem[\protect\citeauthoryear{{Knigge}, {Drake}, {Long}, {Wade}, {Horne} \&
  {Baptista}}{{Knigge} et~al.}{1998}]{KniggeDrakeLong1998}
{Knigge} C.,  {Drake} N.,  {Long} K.~S.,  {Wade} R.~A.,  {Horne} K.,
  {Baptista} R.,  1998, \apj, 499, 429

\bibitem[\protect\citeauthoryear{{K{\"o}rding}, {Falcke} \&
  {Corbel}}{{K{\"o}rding} et~al.}{2006}]{KoerdingFalckeCorbel2005}
{K{\"o}rding} E.,  {Falcke} H.,    {Corbel} S.,  2006, \aap, 456, 439

\bibitem[\protect\citeauthoryear{{K{\"o}rding}, {Fender} \&
  {Migliari}}{{K{\"o}rding} et~al.}{2006}]{KoerdingFenderMigliari2006}
{K{\"o}rding} E.~G.,  {Fender} R.~P.,    {Migliari} S.,  2006, \mnras, 369,
  1451

\bibitem[\protect\citeauthoryear{{K{\"o}rding}, {Jester} \&
  {Fender}}{{K{\"o}rding} et~al.}{2006}]{KoerdingJesterFender2006}
{K{\"o}rding} E.~G.,  {Jester} S.,    {Fender} R.,  2006, \mnras, 372, 1366

\bibitem[\protect\citeauthoryear{{Linares}, {van der Klis} \&
  {Wijnands}}{{Linares} et~al.}{2006}]{LinaresWijnands2006}
{Linares} M.,  {van der Klis} M.,    {Wijnands} R.,  2006, ArXiv Astrophysics
  e-prints

\bibitem[\protect\citeauthoryear{{Markowitz}, {Edelson}, {Vaughan}, {Uttley},
  {George}, {Griffiths}, {Kaspi}, {Lawrence}, {McHardy}, {Nandra}, {Pounds},
  {Reeves}, {Schurch} \& {Warwick}}{{Markowitz}
  et~al.}{2003}]{MarkowitzEdelsonVaughan2003}
{Markowitz} A.,  {Edelson} R.,  {Vaughan} S.,  {Uttley} P.,  {George} I.~M.,
  {Griffiths} R.~E.,  {Kaspi} S.,  {Lawrence} A.,  {McHardy} I.,  {Nandra} K.,
  {Pounds} K.,  {Reeves} J.,  {Schurch} N.,    {Warwick} R.,  2003, \apj, 593,
  96

\bibitem[\protect\citeauthoryear{{Marsh} \& {Horne}}{{Marsh} \&
  {Horne}}{1998}]{MarshHorne1998}
{Marsh} T.~R.,  {Horne} K.,  1998, \mnras, 299, 921

\bibitem[\protect\citeauthoryear{{Massey}, {Johnson} \&
  {Degioia-Eastwood}}{{Massey} et~al.}{1995}]{MasseyJohnsonDegioiaEastwood1995}
{Massey} P.,  {Johnson} K.~E.,    {Degioia-Eastwood} K.,  1995, \apj, 454, 151

\bibitem[\protect\citeauthoryear{{McClintock} \& {Remillard}}{{McClintock} \&
  {Remillard}}{2006}]{McClintockRemillard2003}
{McClintock} J.,  {Remillard} R.,  2006, in "Compact Stellar X-ray Sources,"
  eds. W.H.G. Lewin and M. v an der Klis, Cambridge University Press

\bibitem[\protect\citeauthoryear{{McHardy}}{{McHardy}}{1988}]{McHardy1988}
{McHardy} I.,  1988, Memorie della Societa Astronomica Italiana, 59, 239

\bibitem[\protect\citeauthoryear{{McHardy}, {Koerding}, {Knigge}, {Uttley} \&
  {Fender}}{{McHardy} et~al.}{2006}]{McHardyKoerding2006}
{McHardy} I.~M.,  {Koerding} E.,  {Knigge} C.,  {Uttley} P.,    {Fender} R.~P.,
   2006, \nat, 444, 730

\bibitem[\protect\citeauthoryear{{Merloni}, {Heinz} \& {Di Matteo}}{{Merloni}
  et~al.}{2003}]{MerloniHeinzdiMatteo2003}
{Merloni} A.,  {Heinz} S.,    {Di Matteo} T.,  2003, \mnras, 345, 1057

\bibitem[\protect\citeauthoryear{{Migliari} \& {Fender}}{{Migliari} \&
  {Fender}}{2006}]{MigliariFender2005b}
{Migliari} S.,  {Fender} R.~P.,  2006, \mnras, 366, 79

\bibitem[\protect\citeauthoryear{{Migliari}, {Fender} \& {van der
  Klis}}{{Migliari} et~al.}{2005}]{MigliariFender2005}
{Migliari} S.,  {Fender} R.~P.,    {van der Klis} M.,  2005, \mnras, pp 799--

\bibitem[\protect\citeauthoryear{{Molkov}, {Revnivtsev}, {Lutovinov} \&
  {Sunyaev}}{{Molkov} et~al.}{2005}]{MolkovRevnivtsevLutovinov2005}
{Molkov} S.,  {Revnivtsev} M.,  {Lutovinov} A.,    {Sunyaev} R.,  2005, \aap,
  434, 1069

\bibitem[\protect\citeauthoryear{{Nowak}}{{Nowak}}{1995}]{Nowak1995}
{Nowak} M.~A.,  1995, \pasp, 107, 1207+

\bibitem[\protect\citeauthoryear{{Orosz}, {Groot}, {van der Klis},
  {McClintock}, {Garcia}, {Zhao}, {Jain}, {Bailyn} \& {Remillard}}{{Orosz}
  et~al.}{2002}]{OroszGroot2002}
{Orosz} J.~A.,  {Groot} P.~J.,  {van der Klis} M.,  {McClintock} J.~E.,
  {Garcia} M.~R.,  {Zhao} P.,  {Jain} R.~K.,  {Bailyn} C.~D.,    {Remillard}
  R.~A.,  2002, \apj, 568, 845

\bibitem[\protect\citeauthoryear{{Orosz}, {McClintock}, {Remillard} \&
  {Corbel}}{{Orosz} et~al.}{2004}]{OroszMcClintockRemillard2004}
{Orosz} J.~A.,  {McClintock} J.~E.,  {Remillard} R.~A.,    {Corbel} S.,  2004,
  \apj, 616, 376

\bibitem[\protect\citeauthoryear{{Pandel}, {C{\'o}rdova} \& {Howell}}{{Pandel}
  et~al.}{2003}]{PandelHowell2003}
{Pandel} D.,  {C{\'o}rdova} F.~A.,    {Howell} S.~B.,  2003, \mnras, 346, 1231

\bibitem[\protect\citeauthoryear{{Pottschmidt}, {Wilms}, {Nowak}, {Pooley},
  {Gleissner}, {Heindl}, {Smith}, {Remillard} \& {Staubert}}{{Pottschmidt}
  et~al.}{2003}]{Pottschmidt2002}
{Pottschmidt} K.,  {Wilms} J.,  {Nowak} M.~A.,  {Pooley} G.~G.,  {Gleissner}
  T.,  {Heindl} W.~A.,  {Smith} D.~M.,  {Remillard} R.,    {Staubert} R.,
  2003, \aap, 407, 1039

\bibitem[\protect\citeauthoryear{{Pratt}, {Hassall}, {Naylor}, {Wood} \&
  {Patterson}}{{Pratt} et~al.}{1999}]{PrattHassallNaylor1999}
{Pratt} G.~W.,  {Hassall} B.~J.~M.,  {Naylor} T.,  {Wood} J.~H.,    {Patterson}
  J.,  1999, \mnras, 309, 847

\bibitem[\protect\citeauthoryear{{Psaltis}, {Belloni} \& {van der
  Klis}}{{Psaltis} et~al.}{1999}]{PsaltisBelloni1999}
{Psaltis} D.,  {Belloni} T.,    {van der Klis} M.,  1999, \apj, 520, 262

\bibitem[\protect\citeauthoryear{{Ritter} \& {Kolb}}{{Ritter} \&
  {Kolb}}{2003}]{RitterKolb2003}
{Ritter} H.,  {Kolb} U.,  2003, \aap, 404, 301

\bibitem[\protect\citeauthoryear{{Shahbaz}, {Fender} \& {Charles}}{{Shahbaz}
  et~al.}{2001}]{ShahbazFenderCharles2001}
{Shahbaz} T.,  {Fender} R.,    {Charles} P.~A.,  2001, \aap, 376, L17

\bibitem[\protect\citeauthoryear{{Shaw}, {Mowlavi}, {Rodriguez}, {Ubertini},
  {Capitanio}, {Ebisawa}, {Eckert}, {Courvoisier}, {Produit}, {Walter} \&
  {Falanga}}{{Shaw} et~al.}{2005}]{ShawMowlaviRodriguez2005}
{Shaw} S.~E.,  {Mowlavi} N.,  {Rodriguez} J.,  {Ubertini} P.,  {Capitanio} F.,
  {Ebisawa} K.,  {Eckert} D.,  {Courvoisier} T.~J.-L.,  {Produit} N.,  {Walter}
  R.,    {Falanga} M.,  2005, \aap, 432, L13

\bibitem[\protect\citeauthoryear{{Sion}, {Cheng}, {Sparks}, {Szkody}, {Huang}
  \& {Hubeny}}{{Sion} et~al.}{1997}]{SionChengSparks1997}
{Sion} E.~M.,  {Cheng} F.~H.,  {Sparks} W.~M.,  {Szkody} P.,  {Huang} M.,
  {Hubeny} I.,  1997, \apjl, 480, L17

\bibitem[\protect\citeauthoryear{{Sobczak}, {McClintock}, {Remillard}, {Cui},
  {Levine}, {Morgan}, {Orosz} \& {Bailyn}}{{Sobczak}
  et~al.}{2000}]{SobczakMcClintockRemillard2000}
{Sobczak} G.~J.,  {McClintock} J.~E.,  {Remillard} R.~A.,  {Cui} W.,  {Levine}
  A.~M.,  {Morgan} E.~H.,  {Orosz} J.~A.,    {Bailyn} C.~D.,  2000, \apj, 544,
  993

\bibitem[\protect\citeauthoryear{{Suleimanov}, {Neustroev}, {Borisov} \&
  {Fioktistova}}{{Suleimanov} et~al.}{2004}]{SuleimanovNeustroevBorisov2004}
{Suleimanov} V.~F.,  {Neustroev} V.~V.,  {Borisov} N.~V.,    {Fioktistova}
  I.~S.,  2004, in {Tovmassian} G.,  {Sion} E.,  eds, Revista Mexicana de
  Astronomia y Astrofisica Conference Series {Time-resolved spectroscopy and
  Doppler tomography of UX UMa}.
pp 270--270

\bibitem[\protect\citeauthoryear{{Thompson}, {Rothschild}, {Tomsick} \&
  {Marshall}}{{Thompson} et~al.}{2005}]{ThompsonRothschildTomsick2005}
{Thompson} T.~W.~J.,  {Rothschild} R.~E.,  {Tomsick} J.~A.,    {Marshall}
  H.~L.,  2005, \apj, 634, 1261

\bibitem[\protect\citeauthoryear{{Tomsick}, {Kalemci}, {Corbel} \&
  {Kaaret}}{{Tomsick} et~al.}{2003}]{TomsickKalemciCorbel2003}
{Tomsick} J.~A.,  {Kalemci} E.,  {Corbel} S.,    {Kaaret} P.,  2003, \apj, 592,
  1100

\bibitem[\protect\citeauthoryear{{Trudolyubov}}{{Trudolyubov}}{2001}]{Trudolyu%
bov2001}
{Trudolyubov} S.~P.,  2001, \apj, 558, 276

\bibitem[\protect\citeauthoryear{{Uttley} \& {McHardy}}{{Uttley} \&
  {McHardy}}{2005}]{UttleyMcHardy2005}
{Uttley} P.,  {McHardy} I.~M.,  2005, \mnras, pp 818--

\bibitem[\protect\citeauthoryear{{Uttley}, {McHardy} \& {Papadakis}}{{Uttley}
  et~al.}{2002}]{UttleyMcHardyPapadakis2002}
{Uttley} P.,  {McHardy} I.~M.,    {Papadakis} I.~E.,  2002, \mnras, 332, 231

\bibitem[\protect\citeauthoryear{{van der Klis}}{{van der
  Klis}}{2001}]{Klis2001}
{van der Klis} M.,  2001, \apj, 561, 943

\bibitem[\protect\citeauthoryear{{van Straaten}, {van der Klis} \&
  {M{\'e}ndez}}{{van Straaten} et~al.}{2003}]{vanStraaten2003}
{van Straaten} S.,  {van der Klis} M.,    {M{\'e}ndez} M.,  2003, \apj, 596,
  1155

\bibitem[\protect\citeauthoryear{{Vignarca}, {Migliari}, {Belloni}, {Psaltis}
  \& {van der Klis}}{{Vignarca} et~al.}{2003}]{VignarcaMigliariBelloni2003}
{Vignarca} F.,  {Migliari} S.,  {Belloni} T.,  {Psaltis} D.,    {van der Klis}
  M.,  2003, \aap, 397, 729

\bibitem[\protect\citeauthoryear{{Warner} \& {Woudt}}{{Warner} \&
  {Woudt}}{2006}]{WarnerWoudt2006}
{Warner} B.,  {Woudt} P.~A.,  2006, \mnras, 367, 1562

\bibitem[\protect\citeauthoryear{{Warner}, {Woudt} \& {Pretorius}}{{Warner}
  et~al.}{2003}]{WarnerWoudtPretorius2003}
{Warner} B.,  {Woudt} P.~A.,    {Pretorius} M.~L.,  2003, \mnras, 344, 1193

\bibitem[\protect\citeauthoryear{{Wijnands}, {Mendez}, {van der Klis},
  {Psaltis}, {Kuulkers} \& {Lamb}}{{Wijnands}
  et~al.}{1998}]{WijnandsMendezPsaltis1998}
{Wijnands} R.,  {Mendez} M.,  {van der Klis} M.,  {Psaltis} D.,  {Kuulkers} E.,
     {Lamb} F.~K.,  1998, \apjl, 504, L35

\bibitem[\protect\citeauthoryear{{Wilms}, {Nowak}, {Pottschmidt}, {Pooley} \&
  {Fritz}}{{Wilms} et~al.}{2006}]{WilmsNowakPottschmidt2006}
{Wilms} J.,  {Nowak} M.~A.,  {Pottschmidt} K.,  {Pooley} G.~G.,    {Fritz} S.,
  2006, \aap, 447, 245

\bibitem[\protect\citeauthoryear{{Wood}, {Horne}, {Berriman} \& {Wade}}{{Wood}
  et~al.}{1989}]{WoodHorneBerriman1989}
{Wood} J.~H.,  {Horne} K.,  {Berriman} G.,    {Wade} R.~A.,  1989, \apj, 341,
  974

\bibitem[\protect\citeauthoryear{{Woudt} \& {Warner}}{{Woudt} \&
  {Warner}}{2002}]{WoudtWarner2002}
{Woudt} P.~A.,  {Warner} B.,  2002, \mnras, 333, 411

\bibitem[\protect\citeauthoryear{{Zdziarski}, {Gierli{\'n}ski}, {Rao},
  {Vadawale} \& {Miko{\l}ajewska}}{{Zdziarski} et~al.}{2005}]{ZdziarskiRao2005}
{Zdziarski} A.~A.,  {Gierli{\'n}ski} M.,  {Rao} A.~R.,  {Vadawale} S.~V.,
  {Miko{\l}ajewska} J.,  2005, \mnras, 360, 825

\end{thebibliography}

\end{document}